\documentclass[aps,prl,10pt,superscriptaddress,floatfix,twocolumn,notitlepage]{revtex4-1}
\usepackage{graphicx,graphics}
\usepackage{amsmath,amssymb,amsfonts}
\usepackage{bm}
\usepackage{color}
\usepackage{ulem}
\usepackage[abs]{overpic}
\usepackage[breaklinks=true,colorlinks,citecolor=blue,linkcolor=blue,urlcolor=blue]{hyperref}
\begin{document}
\author{Iacopo Torre}
\email{iacopo.torre@icfo.eu}
\affiliation{ICFO-Institut de Ci\`{e}ncies Fot\`{o}niques, The Barcelona Institute of Science and Technology, Av. Carl Friedrich Gauss 3, 08860 Castelldefels (Barcelona),~Spain}
\author{Luan Vieira de Castro}
\affiliation{Department of Physics, University of Antwerp, Groenenborgerlaan 171, B-2020 Antwerpen,~Belgium}
\affiliation{Departamento de F\'{i}sica, Universidade Federal do Cear\'{a}, Caixa Postal 6030, Campus do Pici, Fortaleza, Cear\'{a} 60455-900,~Brazil}
\author{Ben Van Duppen}
\affiliation{Department of Physics, University of Antwerp, Groenenborgerlaan 171, B-2020 Antwerpen,~Belgium}
\author{David Barcons Ruiz}
\affiliation{ICFO-Institut de Ci\`{e}ncies Fot\`{o}niques, The Barcelona Institute of Science and Technology, Av. Carl Friedrich Gauss 3, 08860 Castelldefels (Barcelona),~Spain}
\author{Fran\c{c}ois M. Peeters}
\affiliation{Department of Physics, University of Antwerp, Groenenborgerlaan 171, B-2020 Antwerpen,~Belgium}
\author{Frank H.L. Koppens}
\affiliation{ICFO-Institut de Ci\`{e}ncies Fot\`{o}niques, The Barcelona Institute of Science and Technology, Av. Carl Friedrich Gauss 3, 08860 Castelldefels (Barcelona),~Spain}
\affiliation{ICREA-Instituci\'{o} Catalana de Recerca i Estudis Avan\c{c}ats, Passeig de Llu\'{i}s Companys 23, 08010 Barcelona,~Spain}
\author{Marco Polini}
\affiliation{Istituto Italiano di Tecnologia, Graphene Labs, Via Morego 30, I-16163 Genova,~Italy}
\affiliation{School of Physics \& Astronomy, University of Manchester, Oxford Road, Manchester M13 9PL,~United Kingdom}
\title{Acoustic plasmons at the crossover between the collisionless and hydrodynamic regimes in two-dimensional electron liquids}
\begin{abstract}
Hydrodynamic flow in two-dimensional electron systems has so far been probed only by dc transport and scanning gate microscopy measurements. In this work we discuss theoretically signatures of the hydrodynamic regime in near-field optical microscopy. We analyze the dispersion of acoustic plasmon modes in two-dimensional electron liquids using a non-local conductivity that takes into account the effects of (momentum-conserving) electron-electron collisions, (momentum-relaxing) electron-phonon and electron-impurity collisions, and many-body interactions beyond the celebrated Random Phase Approximation. We derive the dispersion and, most importantly, the damping of acoustic plasmon modes and their coupling to a near-field probe, identifying key experimental signatures of the crossover between collisionless and hydrodynamic regimes.
\end{abstract}
\maketitle

{\it Introduction.}---In electron systems a collective charge mode exists at frequency above the threshold for intra-band electron-hole excitations. This mode is called ``plasmon''~\cite{giuliani_vignale_book,pines_nozieres_book} and is particularly useful for technological applications in the case of two-dimensional (2D) electron systems. In this case indeed plasmons are gapless modes typically falling in the mid-infrared~\cite{fei_nature_2012,chen_nature_2012,woessner_natmat_2015} or Terahertz (THz)~\cite{ju_nat_nano_2011,alonso_gonzalez_natnano_2017,lundeberg_science_2017} frequency ranges.

\begin{figure}[h!]
\includegraphics[scale=.8]{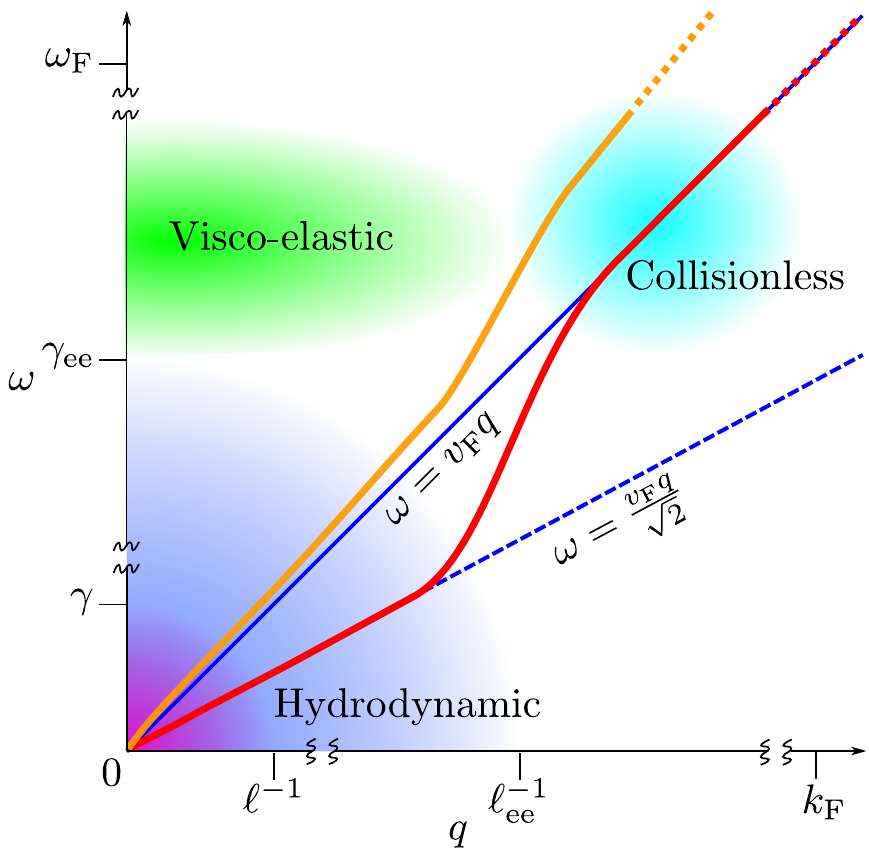}
\caption{\label{fig:sketch} (Color online)
Sketch of the $q$-$\omega$ plane showing the relevant frequency and length scales for the problem at hand, and the plasmon dispersion (red and orange lines) for two different values of the screening parameter $\Lambda$ defined in Eq.~(\ref{eq:screening_parameter}).
Red line: $\Lambda \gg 1$. Orange line: $\Lambda<1$.
The blue solid line is the electron dispersion $\omega=v_{\rm F} q$ while the blue dashed line is the sound dispersion $\omega=v_{\rm F} q/\sqrt{2}$ (ignoring here many-body corrections). Different regimes of linear response are highlighted. In the hydrodynamic regime (blue shaded region) the Navier-Stokes equation (\ref{eq:navierstokes}) is applicable. In the overdamped regime (magenta shaded region) Eq.~(\ref{eq:navierstokes}) is still applicable but plasmons are strongly damped. In the visco-elastic regime (green shaded region) Eq.~(\ref{eq:navierstokes}) can still be applied considering a frequency-dependent complex viscosity~\cite{conti_prb_1999,pellegrino_prb_2017}.
}
\end{figure}

In recent years plasmons in 2D materials~\cite{grigorenko_natphot_2012,basov_science_2016,low_natmat_2017} such as graphene have attracted a great deal of attention because of their ability to confine light on length scales much shorter than the free-space wavelength~\cite{lundeberg_science_2017,alcaraz_science_2018}, their long lifetimes~\cite{woessner_natmat_2015,ni_nature_2018}, and their gate tunability~\cite{yan_nat_nano_2012,fei_nature_2012,chen_nature_2012,woessner_natmat_2015}.

Due to the long-range nature of the bare electron-electron (e-e) interaction, plasmons in 2D materials on a dielectric substrate have a long-wavelength ``unscreened'' dispersion of the form~\cite{giuliani_vignale_book,pines_nozieres_book} $\omega \propto \sqrt{q}$, where $\omega$ is the angular frequency and $q$ is the in-plane wave vector.
Conversely, if the long-range part of the e-e interaction is screened by e.g. a nearby conducting gate, the plasmon dispersion is modified into an acoustic one (see e.g.~Ref.~\onlinecite{santoro_prb_1988}), $\omega \propto q$.

Acoustic plasmons (APs)~\cite{alonso_gonzalez_natnano_2017,lundeberg_science_2017,santoro_prb_1988,principi_ssc_2011,principi_prb_2018} are particularly interesting because they can achieve larger mode confinement with respect to their unscreened counterpart. 
This happens for two reasons. First, an AP is more confined in the vertical direction due to the presence of the metallic gate~\cite{alcaraz_science_2018}, with the largest part of the electromagnetic energy density being localized between the gate and the 2D material. 
Second, due to the screening of the long-range part of the Coulomb interaction, APs are softer (because the restoring force is reduced) and carry high values of $q$, for a given value of $\omega$. This allows the study of interesting quantum non-local effects~\cite{lundeberg_science_2017}, which become important when the plasmon dispersion gets close to the boundary of the intra-band electron-hole continuum located at $\omega=v_{\rm F}^* q$, $v_{\rm F}^*$ being the quasiparticle velocity. With the term ``quasiparticle'' velocity we mean the Fermi velocity as dressed by electron-electron (e-e) interactions~\cite{giuliani_vignale_book,pines_nozieres_book,kotov_rmp_2012}. The same jargon and notation will be used below for the Drude weight ${\cal D}^{*}$, the density-of-states at the Fermi energy ${\cal N}^{*}$, etc. The same quantities without the ``$*$'' symbol, e.g.~$v_{\rm F}$, ${\cal D}$, ${\cal N}$, etc,  will denote instead the non-interacting counterparts.

In 2D conducting materials of extremely high electronic quality, such as graphene encapsulated in hexagonal Boron Nitride~\cite{wang_science_2013}, e-e interactions induce, in the intermediate-to-high-temperature regime, the so-called hydrodynamic transport regime.
In this regime, e-e collisions are so frequent that they can establish a local thermal quasi-equilibrium.
This happens when the e-e mean-free-path $\ell_{\rm ee}\equiv v_{\rm F}^* \tau_{\rm ee}$ (here $\tau_{\rm ee}$ is the e-e scattering time~\cite{giuliani_prb_1982,quian_prb_2005,li_prb_2013,polini_normale_2016,principi_prb_2016}) is much shorter than both the mean-free-path for momentum-relaxing collisions with phonons or impurities $\ell \equiv v_{\rm F}^* \tau$ and the characteristic wavelength~\cite{torre_prb_2015,bandurin_science_2016} $1/q$ of external perturbations.
In the ac regime, we should also require~\cite{torre_prb_2015,bandurin_science_2016,pellegrino_prb_2017} the angular frequency of the perturbation $\omega$ to be much smaller than the e-e scattering rate $1/\tau_{\rm ee}$. Transport signatures of hydrodynamic behaviour have been found in different high-quality materials like single- and bi-layer graphene~\cite{bandurin_science_2016,krishna_kumar_natphys_2017,crossno_science_2016,bandurin_natcomm_2018}, GaAs/AlGaAs heterostructures~\cite{dejong_prb_1995,braem_arxiv_2018}, and ${\rm PdCoO}_2$~\cite{moll_science_2016}. 

The rate $\gamma \equiv 1/\tau$ of momentum non-conserving collisions with impurities and phonons and the e-e scattering rate $\gamma_{\rm ee}\equiv 1/\tau_{\rm ee}$ define several regimes in the $q$-$\omega$ plane, which are sketched in Fig.~\ref{fig:sketch}.  

In the hydrodynamic regime~\cite{landaufluidmechanics} and at the level of linear-response theory, the electron liquid can be described by the continuity equation $i\omega n(\bm r,\omega)=\nabla \cdot \bm J(\bm r,\omega)$, $n(\bm r,\omega)$ being the deviation of the particle density from its equilibrium value $\bar{n}$ and $\bm J(\bm r,\omega)$ the particle current, and the Navier-Stokes equation~\cite{torre_prb_2015,bandurin_science_2016,pellegrino_prb_2017}
\begin{eqnarray}\label{eq:navierstokes} 
&-&i\omega \bm J(\bm r,\omega)  = 
 -\gamma \bm J(\bm r,\omega)+\nu^{*}\nabla^2\bm J(\bm r,\omega)+\nonumber\\
& -&\frac{\mathcal{D}^*}{\mathcal{D}}\left[ \frac{e\bar{n}}{m} {\bm E}(\bm r,\omega)+\frac{1}{\bar{n}mK^*}\nabla n(\bm r, \omega)\right]~.
\end{eqnarray}
Here, ${\bm E}(\bm r,\omega)$ is the electric field, $e$ is the elementary charge, $m \equiv \hbar k_{\rm F}/v_{\rm F}$ is the bare effective mass, $k_{\rm F}$ being the Fermi wave vector, $K^{*}=[\bar{n}\partial P/\partial \bar{n}]^{-1}$ is the compressibility~\cite{pines_nozieres_book,giuliani_vignale_book,asgari_adp_2014}, $P=P(\bar{n})$ being the pressure, $\nu^{*}$ is the kinematic viscosity~\cite{torre_prb_2015,bandurin_science_2016,pellegrino_prb_2017,landaufluidmechanics}, $\mathcal{D}^*$ ($\mathcal{D}$) is the Drude weight of the interacting~\cite{abedinpour_prb_2011,levitov_prb_2013} (non-interacting) electron system.
A derivation of Eq.~(\ref{eq:navierstokes}) is given in Sect.~I of Ref.~\onlinecite{note_supplementary}.

In this work we identify signatures of the transition between the hydrodynamic ($\omega \ll \gamma_{\rm ee}$) and collisionless ($\omega \gg \gamma_{\rm ee}$) regimes in the dispersion and, most importantly, the damping of AP modes. In the case of single-layer graphene (SLG) at room temperature, for example, $\tau_{\rm ee} \approx 0.15~{\rm ps}$ at typical carrier densities~\cite{principi_prb_2016}  ($\bar{n} = 1.0 \times 10^{12}~{\rm cm}^{-2}$, say) and the crossover is expected to occur in the THz range. Our work is structured as follows. We first introduce the two main ingredients of our theory: the non-local longitudinal conductivity $\sigma_{\rm L}(q,\omega)$---Eq.~(\ref{eq:sigma_final})---and the interaction potential $v_{q,\omega}$, both calculated in the long-wavelength limit.
We then find AP modes, which are described by an equation of the form $q_{\rm p} = q_{\rm p}(\omega)$ for every real frequency $\omega$. Here $q_{\rm p}$ is a complex wave vector $q_{\rm p} = {\rm Re}(q_{\rm p})+ i {\rm Im}(q_{\rm p})$, which gives access to both dispersion and damping. Finally, we analyze the coupling of these  modes to a near-field probe and discuss our results.

{\it The non-local conductivity from Landau kinetic theory.}---The response of a 2D electron liquid to an external scalar potential can be calculated using Landau kinetic equation~\cite{pines_nozieres_book, giuliani_vignale_book} for a normal Fermi liquid, which governs the response of the quasiparticle distribution function to slowly-varying electromagnetic fields~\cite{conti_prb_1999,pellegrino_prb_2017}. Its use is justified when the excitation wavelength is sufficiently long compared to the inverse of the Fermi wave vector $k_{\rm F}$, and when the excitation energy $\hbar\omega$ is sufficiently small compared to the Fermi energy $E_{\rm F}$, and to the energy of the lowest inter-band excitation $E_{\rm g}$. 

As detailed in Sects.~I-II  of Ref.~\onlinecite{note_supplementary}, the linearized kinetic equation can be solved by using a simple ansatz~\cite{pellegrino_prb_2017}. After lengthy but straightforward algebra, we find the following expression for the longitudinal non-local conductivity~\cite{note_conductivity}, which controls the current response to an electric field parallel to ${\bm q}$:
\begin{widetext}
\begin{equation}\label{eq:sigma_final}
\begin{split}
&\sigma_{\rm L}(q,\omega)
=\dfrac{i \mathcal{D}^*/\pi}{\omega+i\gamma +\dfrac{\omega+i\gamma+i\gamma_{\rm ee}}{2}\dfrac{\mathcal{D}^*}{\mathcal{D}}\dfrac{v_{\rm F}}{v_{\rm F}^*}\left[\sqrt{1-\left(\dfrac{v_{\rm F}^*q}{\omega+i\gamma+i\gamma_{\rm ee}}\right)^2}-1\right]- \dfrac{1}{2}\dfrac{\mathcal{D}^*}{\mathcal{D}}\dfrac{K}{K^*}\dfrac{v_{\rm F}^2q^2}{\omega}}.
\end{split}
\end{equation}
\end{widetext}
Here, $\mathcal{D}=\pi e^2 \bar{n}/m$ ($K= {\cal N}/\bar{n}^2$) is the Drude weight (compressibility) of the non-interacting system, ${\cal N}= N_{\rm f} m/(2\pi \hbar^2)$ being the density-of-states at the Fermi energy and $N_{\rm f}$  the number of fermion flavors (e.g.~$N_{\rm f}=4$ for graphene).
In Landau theory of Fermi liquids~\cite{pines_nozieres_book,giuliani_vignale_book}, $K/K^* = (v_{\rm F}^{*}/v_{\rm F})(1+ F^{\rm s}_{0})$ and $\mathcal{D}^*/\mathcal{D}=(v_{\rm F}^{*}/v_{\rm F})(1+F_1^{\rm s})$, where $F^{\rm s}_{0(1)}$ is the spin-symmetric dimensionless Landau parameter in the $s$ ($p$) angular momentum channel~\cite{pines_nozieres_book,giuliani_vignale_book,note_landau_parameters}.
The many-body corrections $v_{\rm F}^*/v_{\rm F}$, $K^{*}/K$, and $\mathcal{D}^*/\mathcal{D}$ can be calculated from approximated theories~\cite{kotov_rmp_2012,asgari_adp_2014,abedinpour_prb_2011,lundeberg_science_2017} and are fundamental for a quantitative interpretation of experimental data since, for example, $v_{\rm F}^*/v_{\rm F}\approx 1.3$~\cite{kotov_rmp_2012}, $K^*/K\approx 0.8$~\cite{lundeberg_science_2017}, and $\mathcal{D}^*/\mathcal{D}\approx 1.5$~\cite{abedinpour_prb_2011} in SLG at densities on the order of~$10^{12}~{\rm cm}^{-2}$.  

In deriving Eq.~(\ref{eq:sigma_final}) we made the following assumptions.
i) The momentum-conserving and the momentum-relaxing collisions are described by one parameter each, i.e.~differences between the relaxation times of the different angular components of the distribution function~\cite{ledwith_arxiv_2017} and the difference between $\tau_{\rm ee}$ and the viscosity time $\tau_{\rm v}$~\cite{principi_prb_2016} are neglected.
ii) Only the zeroth- and first-order, spin symmetric, Landau parameters $F^{\rm s}_{0(1)}$ are considered. Higher-angular-momentum Landau parameters $F^{\rm s}_{l}$ with $l\geq 2 $ are typically smaller, unless the system is highly correlated. We used these assumptions to derive the simplest yet highly-non-trivial model for the non-local longitudinal conductivity. However, the technique we used in our derivation, based on analytical inversion of tridiagonal matrices~\cite{lorentzen_book,note_supplementary}, easily allows the introduction of different scattering rates for the different harmonics of the distribution function~\cite{ledwith_arxiv_2017} as well as higher-order Landau parameters.  
\begin{figure}
\begin{overpic}[scale=1]{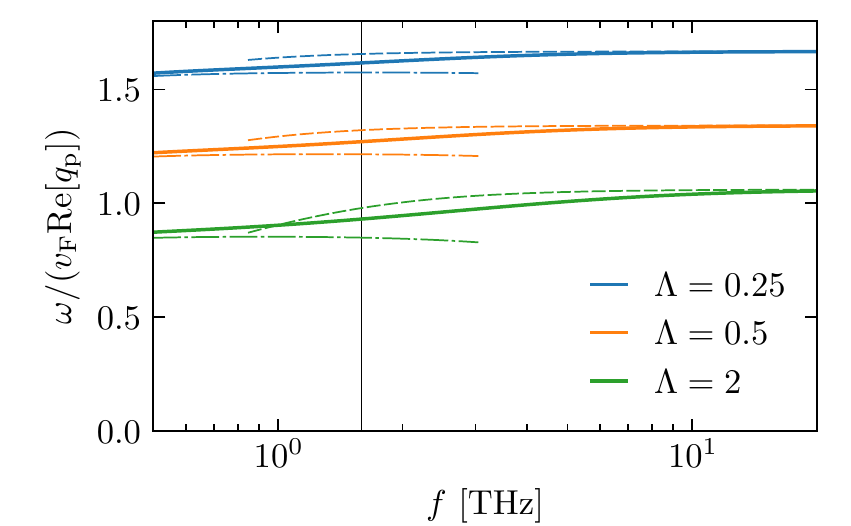}\put(2,145){(a)}
\end{overpic}\\
\begin{overpic}[scale=1]{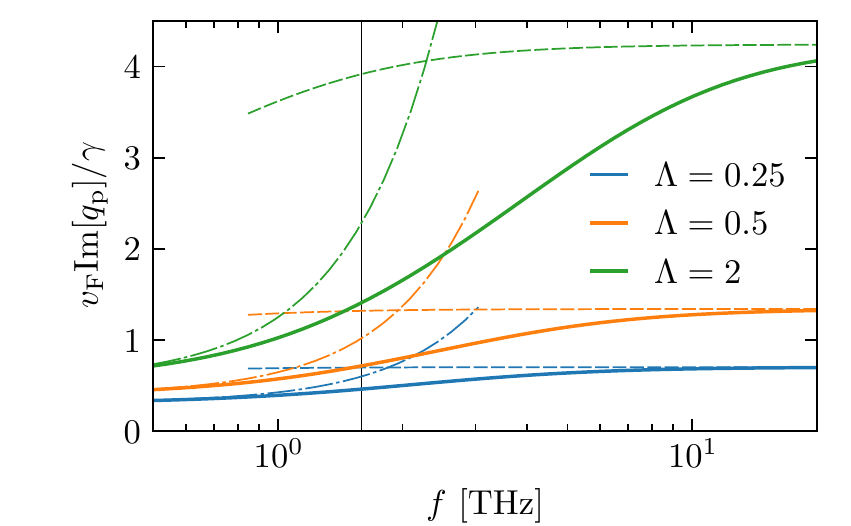}\put(2,145){(b)}
\end{overpic}\\
\caption{\label{fig:realomega}(Color online) AP phase velocity normalized to the Fermi velocity (a), and AP damping, normalized to the extrinsic damping $\gamma$, (b), as functions of the frequency $f=\omega/(2\pi)$, for different values of the screening parameter: $\Lambda=0.25$ (blue), $\Lambda=0.5$ (orange), and $\Lambda=2$ (green). 
Results in this figure have been obtained by setting $\gamma=10^{12}~{\rm s}^{-1}$, $\gamma_{\rm ee}=10^{13}~{\rm s}^{-1}$, and neglecting, for the sake of simplicity, many-body renormalizations by setting $v_{\rm F}^*/v_{\rm F}=K^*/K=\mathcal{D}^*/\mathcal{D}=1$.
For each value of $\Lambda$, the solid line denotes the result of the solution of $\epsilon_{\rm L}(q,\omega)=0$, while the dashed (dash-dotted) line represents the asymptotic collisionless (hydrodynamic) result. The vertical black lines mark the frequency $2\pi f=\gamma_{\rm ee}$ around which the crossover occurs}.
\end{figure}
Eq.~(\ref{eq:sigma_final}) is the first important result of this work because, despite its simplicity, it i) embodies a wealth of physical effects, including many-body effects beyond the Random Phase Approximation (RPA), ii) allows us to span the whole frequency range, from the hydrodynamic to the collisionless regime, and iii) is valid with no assumptions on the relative values of the parameters, other than the ones mentioned previously for the applicability of Landau kinetic equation. In what follows we will anyway assume that $\gamma_{\rm ee}\gg \gamma$ because the hydrodynamic regime is relevant only in this case.

We now look at four special limits of Eq.~(\ref{eq:sigma_final}). 
i) We first set $q=0$, i.e.~we consider the local conductivity. In this case, Eq.~(\ref{eq:sigma_final}) reduces to a Drude-like formula with a renormalized Drude weight $\mathcal{D}^*$ and a damping rate $\gamma$ induced solely by momentum-non-conserving collisions. The e-e collision rate $\gamma_{\rm ee}$ appears at order $q^2$. Note that e-e interactions fully disappear from $\sigma_{\rm L}(0,\omega)$ in a Galilean invariant electron system where ${\cal D}^{*} = {\cal D}$ because in this case~
\cite{pines_nozieres_book,giuliani_vignale_book} $v^{*}_{\rm F}/v_{\rm F} = 1/(1+F^{\rm s}_{1})$.
ii) Second, expanding to second order in $|v_{\rm F}^*q/(\omega+i\gamma+i\gamma_{\rm ee})|$ the square root in the denominator of Eq.~(\ref{eq:sigma_final}) and taking the limit $\omega \ll \gamma_{\rm ee}$, we obtain the hydrodynamic non-local conductivity~\cite{principi_prb_2016}
\begin{equation}\label{eq:sigmahydro}
\sigma_{\rm L}^{\rm h}(q,\omega)=\dfrac{i \mathcal D^*/\pi}{\omega+i\gamma +q^2\left(i\nu^*-\dfrac{\mathcal{D}^*}{\bar{n} m \mathcal{D}K^{*}\omega}\right)}~,
\end{equation}
where $\nu^* \equiv \mathcal{D}^*v_{\rm F}^*v_{\rm F}/[4\mathcal{D}(\gamma_{\rm ee}+\gamma)]$. Ignoring many-body renormalizations, our result for $\nu^{*}$ reduces to the ``classical'' formula for the viscosity of an electron gas~\cite{steinberg_pr_1958,pellegrino_prb_2017}, while for Galileian invariant systems it reduces to the expression given in Ref.~\cite{conti_prb_1999} with $F_2^{\rm s}=0$. 
The quantity $\sigma_{\rm L}^{\rm h}(q,\omega)$ can be obtained directly by using Eq.~(\ref{eq:navierstokes}) coupled to the continuity equation. iii) Third, if both many-body renormalizations and e-e collisions are neglected we recover the response function used in Ref.~\cite{giuliani_prb_1984} to discuss the effect of diffusion (i.e.~electron-impurity collisions) on 2D unscreened plasmons.
iv) Finally, if the scattering rates $\gamma$ and $\gamma_{\rm ee}$ are both sent to zero, the long-wavelength ($q\ll k_{\rm F}$) limit of the collisionless conductivity of a 2D electron system~\cite{stern_prl_1967} with parameters renormalized by e-e interactions is recovered.

\begin{figure}
\includegraphics[scale=1]{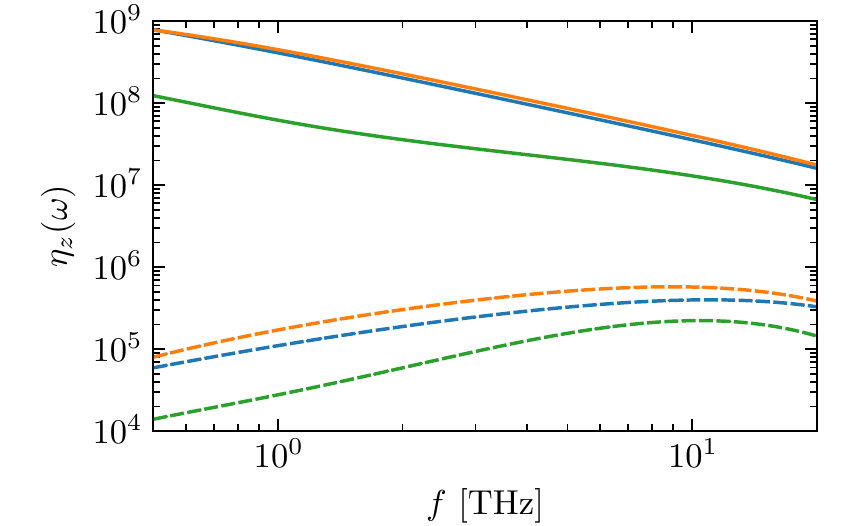}
\caption{\label{fig:coupling} (Color online)
Coupling efficiency $\eta_{z}(\omega)$ as a function of frequency. Results in this figure refer to SLG separated from a metal gate by an hBN spacer of thickness $d=4~{\rm nm}$, having ${\bar \epsilon}_{xx}={\bar \epsilon}_{yy}=6.68$ and ${\bar \epsilon}_{zz}=3.56$. 
Solid lines correspond to excitation in the center of the spacer $z=-2~{\rm nm}$, while dashed curves correspond to $z=10~{\rm nm}$, above SLG.
Different colors refer to different values of the screening parameter: $\Lambda=0.25$ (blue), $\Lambda=0.5$ (orange), and $\Lambda=2$ (green). All other parameters are as in Fig~\ref{fig:realomega}.}
\end{figure}

{\it The screened e-e interaction.}---The dispersion of plasmons in a material depends also on the interaction potential $v_{q,\omega}$ between charges in the material itself.
This quantity relates the Fourier transform of $n(\bm q,\omega)$ to the Fourier transform of the induced (i.e.~Hartree) scalar potential $V_{\rm ind}(\bm q,\omega)$, i.e.~$V_{\rm ind}(\bm q,\omega)=v_{q,\omega}n(\bm q,\omega)$. In 2D materials the interaction potential is strongly affected by the presence of nearby dielectrics or conductors.
The interaction potential for generic layered structures can be easily calculated~\cite{tomadin_prl_2015,alonso_gonzalez_natnano_2017}. For example, for a graphene sheet encapsulated between hBN slabs of different thickness and in the presence of a metallic gate has been calculated in Ref.~\onlinecite{alonso_gonzalez_natnano_2017}. For low frequencies (i.e.~low compared to all, e.g.~phonon, features in the dielectric functions of the nearby dielectrics) and long wavelengths (i.e.~for $q$ much smaller than the inverse of the dielectric thickness) $v_{q,\omega}$ can be safely replaced by its limit 
$v_{q,\omega}\approx \lim_{q,\omega\to 0}v_{q,\omega} \equiv e^2/C$, $C$ being the capacitance per unit area of the structure (see Sect.~III of Ref.~\onlinecite{note_supplementary}). If we consider a structure made of a perfectly conducting gate parallel to the 2D electron system and separated along the $\hat{\bm z}$-direction by a dielectric spacer of thickness $d$ and dielectric tensor $\bar{\bm \epsilon}$, the capacitance per unit area is $C=\bar{\epsilon}_{zz}/(4\pi d)$, where ${\bar \epsilon}_{zz}$ denotes the tensor component along the $\hat{\bm z}$ direction. For all realistic experimental geometries~\cite{alonso_gonzalez_natnano_2017,lundeberg_science_2017} using e.g.~graphene encapsulated in hBN, the plasmon wavelength is much longer than the thickness of the whole device and, therefore, the replacement $v_{q,\omega}\to e^2/C$, i.e.~the so-called local capacitance approximation (LCA), is fully justified in the THz regime where the hydrodynamic-ballistic crossover takes place. All results reported in Figs.~\ref{fig:realomega} and~\ref{fig:coupling} refer to SLG encapsulated in hBN.

{\it AP velocity and damping.}---Mathematically, plasmons are zeroes of the longitudinal dielectric function~\cite{pines_nozieres_book,giuliani_vignale_book} $\epsilon_{\rm L}(q,\omega)$ of the 2D electron system, $\epsilon_{\rm L}(q,\omega)=1+iq^2v_{q,\omega}\sigma_{\rm L}(q,\omega)/(e^2\omega)$. Using the LCA the latter becomes
\begin{equation}\label{eq:dielectricfunction}
\epsilon_{\rm L} (q,\omega)=1-\Lambda^{-1}\frac{(-i)\pi  q^2v_{\rm F}^2\sigma_{\rm L}( q, \omega)}{2\omega\mathcal{D}},
\end{equation}
where 
\begin{equation}\label{eq:screening_parameter}
\Lambda=\frac{C}{e^2\mathcal{N}},
\end{equation}
is a dimensionless parameter that characterizes how much the e-e interaction is screened by the nearby dielectric environment. 
Numerical values of this important parameter are given in Sect.~IV of Ref.~\cite{note_supplementary} for graphene.

The plasmon equation $\epsilon_{\rm L}(q,\omega)=0$ with $\epsilon_{\rm L}(q,\omega)$ as in Eq.~(\ref{eq:dielectricfunction}) can be
solved for the plasmon wave vector $q_{\rm p}$. We find  $q_{\rm p}(\omega)=(\omega/S_\omega)\sqrt{1+2i \Gamma_\omega/\omega}$, where $S_\omega$ and $\Gamma_\omega$ are real functions of the frequency representing the velocity and the damping of the mode respectively.
These two functions can be calculated analytically (see Sect.~V of Ref.~\onlinecite{note_supplementary}) and the result is shown in Fig.~\ref{fig:realomega}.
We are now interested in the asymptotic behavior of $S_\omega$ and $\Gamma_\omega$ for $\omega \gg \gamma_{\rm ee}$ (collisionless limit) and $\omega \ll \gamma_{\rm ee}$ (hydrodynamic limit). In the former we find
\begin{widetext}
\begin{align}\label{eq:collisionless_results_1}
S_{\rm c} & =\frac{v_{\rm F}(\Lambda^{-1}+\frac{K}{K^*})}
{\sqrt{\dfrac{(2\mathcal{D}v_{\rm F}+4\mathcal{D}v_{\rm F}^*-2\mathcal{D}^*v_{\rm F})(\Lambda^{-1}+\frac{K}{K^*})-\mathcal{D}^*v_{\rm F}^*}{2\mathcal{D}^*v_{\rm F}^*}\left[1+\sqrt{1-\dfrac{16v_{\rm F}^*\mathcal{D}(v_{\rm F}^*\mathcal{D}-v_{\rm F}\mathcal{D}^*)(\Lambda^{-1}+\frac{K}{K^*})^2}{[(2\mathcal{D}v_{\rm F}+4\mathcal{D}v_{\rm F}^*-2\mathcal{D}^*v_{\rm F})(\Lambda^{-1}+\frac{K}{K^*})-\mathcal{D}^*v_{\rm F}^*]^2}}\right]}}~,\\
\Gamma_{\rm c} & =
\frac{\gamma\dfrac{(\mathcal{D}v_{\rm F}+2\mathcal{D}v_{\rm F}^*-\mathcal{D}^*v_{\rm F})S_{\rm h}^2-2(\mathcal{D}v_{\rm F}^*-\mathcal{D}^*v_{\rm F})S_{\rm c}^2}{v_{\rm F}^{3}\mathcal{D}^*}
+\gamma_{\rm ee }\dfrac{S_{\rm c}^2-S_{\rm h}^2}{v_{\rm F}^{2}}}
{\dfrac{(2\mathcal{D}v_{\rm F}+4\mathcal{D}v_{\rm F}^*-2\mathcal{D}^*v_{\rm F})(\Lambda^{-1}+\frac{K}{K^*})-\mathcal{D}^*v_{\rm F}^*}{2\mathcal{D}v_{\rm F}}\sqrt{1-\dfrac{16v_{\rm F}^*\mathcal{D}(v_{\rm F}^*\mathcal{D}-v_{\rm F}\mathcal{D}^*)(\Lambda^{-1}+\frac{K}{K^*})^2}{[(2\mathcal{D}v_{\rm F}+4\mathcal{D}v_{\rm F}^*-2\mathcal{D}^*v_{\rm F})(\Lambda^{-1}+\frac{K}{K^*})-\mathcal{D}^*v_{\rm F}^*]^2}}}~, \label{eq:collisionless_results_2}
\end{align}
\end{widetext}
while in the latter we find
\begin{align}\label{eq:hydro_result_1}
S_{\rm h} & =v_{\rm F}\sqrt{\dfrac{\mathcal{D}^*(\Lambda^{-1}+\frac{K}{K^*})}{2\mathcal{D}}}~,\\
\Gamma_{\rm h} & =\frac{\gamma}{2}+\frac{\mathcal{D}^*v_{\rm F}v_{\rm F}^*\omega^2}{8\mathcal{D}(\gamma+\gamma_{\rm ee})S_{\rm h}^2}~.\label{eq:hydro_result_2}
\end{align}
Eqs.~(\ref{eq:collisionless_results_1})-(\ref{eq:hydro_result_2}) are the second important result of this work. In particular, Eqs.~(\ref{eq:hydro_result_1})-(\ref{eq:hydro_result_2}) can be obtained by directly solving Eq.~(\ref{eq:dielectricfunction}) with the conductivity given in Eq.~(\ref{eq:sigmahydro}) and ignoring terms of order higher than one in $\omega/\gamma_{\rm ee}$.
 
From these results one can easily understand why achieving high values of the screening parameter $\Lambda$ is of pivotal importance to observe the crossover from the collisionless to the hydrodynamic regime.
Indeed, in the limit $\Lambda\to 0$ we have $S_{\rm h}=S_{\rm c}=v_{\rm F}\sqrt{\mathcal{D}^*/(2\mathcal{D}\Lambda)}$ and $\Gamma_{\rm h}=\Gamma_{\rm c}=\gamma/2$. Therefore, for small values of $\Lambda$ no crossover can be observed as $S_{\rm h}=S_{\rm c}$ and $\Gamma_{\rm h}=\Gamma_{\rm c}$, and the damping of the AP mode is completely controlled by momentum-relaxing collision, with $\gamma_{\rm ee}$ dropping out of the problem.

On the other hand, for $\Lambda \gg 1$ the velocities in the two regimes converge to distinct values. 
The velocity of the AP mode in the collisionless regime tends to a value which is close (ignoring here, for the sake of simplicity, many-body corrections) to the Fermi velocity, $S_{\rm c}\to v_{\rm F}$, while in the hydrodynamic regime it converges to the speed of sound in a neutral Fermi liquid~\cite{phan_arxiv_2013,lucas_prb_2018}, i.e.~ $S_{\rm h}\to v_{\rm F}\sqrt{(\mathcal{D}^*K)/(2\mathcal{D}K^*)}\approx v_{\rm F}/\sqrt{2}$. The situation is even more dramatic for the damping $\Gamma_\omega$.
In the hydrodynamic regime, and for $\Lambda \gg1$, we have $\Gamma_{\rm h}\approx \gamma/2+\omega^2/[4(\gamma+\gamma_{\rm ee})]$, while $\Gamma_{\rm c}\approx \gamma+\gamma_{\rm ee}$, implying that the extrinsic dissipation controlled by $\gamma$ becomes twice more efficient with respect to the $\Lambda \ll 1$ case and a new damping mechanism controlled by $\gamma_{\rm ee}$ kicks in.  
In Fig.~\ref{fig:realomega} we show the impact of $\Lambda$ on the real and imaginary parts of $q_{\rm p}$. When frequency increases, the damping starts to acquire a significant contribution from e-e collisions. This shows up as viscous dissipation in the hydrodynamic regime---see the second term in Eq.~(\ref{eq:hydro_result_2}).
In this regime, indeed, the contribution to the damping is proportional to $q^2$ and therefore to $\omega^2$, since we are probing the damping along the AP dispersion.
When frequency is further increased above $\gamma_{\rm ee}$, the e-e contribution to the damping saturates to a finite value. Note that since in hydrodynamic electron liquids $\gamma_{\rm ee}\gg \gamma$, this contribution can be the dominant one even with moderate values of $\Lambda$ and lead to a significant increase of the imaginary part of $q$, as shown in Fig.~\ref{fig:realomega}(b).

{\it Coupling efficiency to a near-field probe.}---In order to design experiments that are able to probe the collisionless to hydrodynamic crossover with light, it is important also to consider the coupling strength of APs to an external field.
We characterize the coupling to an external near-field probe using the quantity $\eta_{z}(\omega)$ defined by the ratio between the power fed into the AP mode by a dipole source of strength $p$ and frequency $\omega$, located at an height $z$, with its axis perpendicular to the 2D liquid, and the power radiated by the same source in vacuum, given by Larmor's formula (see Sect.~VI of Ref.~\cite{note_supplementary}).

In Fig.~\ref{fig:coupling} we show the numerically-calculated dependence of $\eta_{z}(\omega)$ on frequency for different vertical positions $z$ of the dipole for the aforementioned case of a 2D material separated from a perfect metal located at $z=-d$ by a dielectric spacer. 
For long wavelengths, and assuming small dissipation we can approximate $\eta_{z}(\omega)$ as
\begin{equation}
\eta_{z}(\omega)\approx \frac{3\pi |Z|c^3[{\rm Re} (q_{\rm p})]^3}{\bar{\epsilon}\omega^3}\times
\begin{cases}
d{\rm Re}(q_{\rm p})e^{-2{\rm Re}(q_{\rm p}) z}\; z>0\\
[d{\rm Re}(q_{\rm p})]^{-1}\; 0>z>-d,
\end{cases}
\end{equation}
where $Z\equiv [{\rm Re}(q_{\rm p}) \partial_q\epsilon_{\rm L}(q,\omega)|_{q=q_{\rm p}}]^{-1}=-\{2+q_{\rm p} \partial_q\log[\sigma_{\rm L}(q_{\rm p},\omega)]\}^{-1}\approx -1/2$.
Since $q_{\rm p} d$ is a small number, we see that the AP modes are much more coupled to a dipole located between the material and the gate.
This happens because the electric field of AP modes is mainly concentrated in the spacer region~\cite{woessner_acsphot_2017}.
This suggest that to couple efficiently to these modes, structures specially designed for launching plasmons should be put in the region where the field is concentrated.

In summary, we have studied the dispersion and damping of APs in a 2D electron liquid at the crossover between the hydrodynamic and collisionless regimes.
We have found that, in the presence of strong screening by an external gate, both the velocity and the damping of AP modes are enhanced in the collisionless regime, with the enhancement being more dramatic for the damping.
If the screening is strong enough, i.e.~if $\Lambda>1$, well defined APs with a phase velocity smaller than the Fermi velocity $v_{\rm F}$ (but larger than the sound velocity $\approx v_{\rm F}/\sqrt{2}$) are allowed in the hydrodynamic regime.

Notice that some properties of plasmons in 2D Fermi liquids have been discussed in two recent publications, Refs.~\onlinecite{lucas_prb_2018} and~\onlinecite{svintsov_prb_2018}. 
However, the former mainly focusses on the difference between long-range and short-range interactions, and considers only the many-body compressibility renormalization. In the latter work, effects beyond RPA are neglected, and so are momentum non-conserving processes. We have, however, demonstrated that the latter processes are important to correctly describe the plasmon damping and introduce the possibility of having overdamped excitations at low frequencies and long wavelengths, as shown in Eqs.~\eqref{eq:collisionless_results_2} and~\eqref{eq:hydro_result_2}.  
The non-linear electromagnetic response of a Dirac electron fluid at the crossover between the collisionless and hydrodynamic regimes has been discussed in Ref.~\onlinecite{sun_pnas_2018}.

{\it Acknowledgements.}---This work has been sponsored by the European Union's Horizon 2020 research and innovation programme under grant agreement No.~785219---``Graphene Core2'' and via the European Research Council (ERC) grant agreement No. 786285.
B.V.D. is supported by a post-doctoral fellowship of the Flemish Science Foundation (FWO-Vl). F.H.L.K. acknowledges financial support from the Spanish Ministry of Economy and Competitiveness, through the ``Severo Ochoa'' Programme for Centres of Excellence in R\&D (SEV-2015-0522), support by Fundacio Cellex Barcelona, Generalitat de Catalunya through the CERCA program,  and the Mineco grant Plan Nacional (FIS2016-81044-P) and the Agency for Management of University and Research Grants (AGAUR) 2017 SGR 1656.
We thank Niels Hesp and Hanan Hertzig Sheinfux for useful discussions.

\begin{thebibliography}{apsrev}
\bibitem{note_landau_parameters}
Note that we adopted the definition of Landau parameters given in Ref.~\onlinecite{giuliani_vignale_book}. 
We warn the reader that other authors use a definition that differs from ours by a factor of two for the parameters $F^{\rm s}_l$ with $l>0$. 
The definition of $F^{\rm s}_0$ is instead universally accepted.
\end{thebibliography}
\end{document}


\renewcommand{\bibnumfmt}[1]{[S#1]}
\renewcommand{\citenumfont}[1]{S#1}
\renewcommand{\thetable}{S\arabic{table}}
\renewcommand{\theequation}{S\arabic{equation}}
\renewcommand{\thefigure}{S\arabic{figure}}
\author{Iacopo Torre}
\email{iacopo.torre@icfo.eu}
\affiliation{ICFO-Institut de Ci\`{e}ncies Fot\`{o}niques, The Barcelona Institute of Science and Technology, Av. Carl Friedrich Gauss 3, 08860 Castelldefels (Barcelona),~Spain}
\author{Luan Vieira de Castro}
\affiliation{Department of Physics, University of Antwerp, Groenenborgerlaan 171, B-2020 Antwerpen,~Belgium}
\affiliation{Departamento de F\'{i}sica, Universidade Federal do Cear\'{a}, Caixa Postal 6030, Campus do Pici, Fortaleza, Cear\'{a} 60455-900,~Brazil}
\author{Ben Van Duppen}
\affiliation{Department of Physics, University of Antwerp, Groenenborgerlaan 171, B-2020 Antwerpen,~Belgium}
\author{David Barcons Ruiz}
\affiliation{ICFO-Institut de Ci\`{e}ncies Fot\`{o}niques, The Barcelona Institute of Science and Technology, Av. Carl Friedrich Gauss 3, 08860 Castelldefels (Barcelona),~Spain}
\author{Fran\c{c}ois M. Peeters}
\affiliation{Department of Physics, University of Antwerp, Groenenborgerlaan 171, B-2020 Antwerpen,~Belgium}
\author{Frank H.L. Koppens}
\affiliation{ICFO-Institut de Ci\`{e}ncies Fot\`{o}niques, The Barcelona Institute of Science and Technology, Av. Carl Friedrich Gauss 3, 08860 Castelldefels (Barcelona),~Spain}
\affiliation{ICREA-Instituci\'{o} Catalana de Recerca i Estudis Avan\c{c}ats, Barcelona,~Spain}
\author{Marco Polini}
\affiliation{Istituto Italiano di Tecnologia, Graphene Labs, Via Morego 30, I-16163 Genova,~Italy}
\affiliation{School of Physics \& Astronomy, University of Manchester, Oxford Road, Manchester M13 9PL,~United Kingdom}
\title{Supplemental material for ``Acoustic plasmons at the crossover between the collisionless and hydrodynamic regimes in two-dimensional electron liquids''}
\maketitle
\section{Linearized Boltzmann equation}
\label{app:boltzmann}
For sufficiently long-wavelengths (long compared with the inverse of the Fermi wavevector $k_{\rm F}$) and low-frequencies (low with respect to the Fermi energy $E_{\rm F}$ and to the energy $E_{\rm g}$ of the lowest inter-band excitation), an interacting 2D electron system can be described as a gas of weakly-interacting quasiparticles~\cite{pines_nozieres_book,giuliani_vignale_book}.
If the system is in the paramagnetic state and there is no external perturbation coupling to the spin degrees of freedom, the dynamics of quasiparticles is governed by the {\it classical}, spin-independent, Hamiltonian~\cite{pines_nozieres_book,giuliani_vignale_book}
\begin{equation}\label{eq:hamiltonian}
{\cal H}(\bm r, \bm p,t)=\epsilon^*_{\bm p}-e \phi(\bm r)+U_{\rm L}(\bm r,\bm p,t).
\end{equation}
Here, $\epsilon^*_{\bm p}$ is the band energy of an electron with momentum $\bm p$, renormalized by e-e interactions, $\phi(\bm r,t)$ is the electric scalar potential, and $U_{\rm L}(\bm r,\bm p,t)$ is the spin-averaged Landau interaction potential defined by 
\begin{equation}
U_{\rm L}(\bm r,\bm p,t)=\frac{L^2}{2\hbar^2}\sum_{\sigma\sigma'}\int \frac{d{\bm p}'}{(2\pi)^2}f_{\bm p\sigma,\bm p'\sigma'}\delta f^{(1)}(\bm r,\bm p',t),
\end{equation}
where $f_{{\bm p}\sigma,{\bm p'}\sigma'}$ is the Landau interaction function\cite{giuliani_vignale_book} between an electron with momentum $\bm p$ and spin $\sigma$ and an electron with momentum $\bm p'$ and spin $\sigma'$,  $\delta f^{(1)}(\bm r,\bm p,t)$ is the deviation of the one-particle, spin summed, distribution function $f^{(1)}(\bm r,\bm p,t)$ from its equilibrium value, and $L^2$ is the surface of the 2D electron system.
The Landau interaction function describes, in a mean-field way, dynamical exchange and correlation effects arising from the deviation of the occupation numbers of the electronic states from their equilibrium values.

The classical Hamiltonian (\ref{eq:hamiltonian}) determines the response of quasiparticles via the Landau kinetic equation~\cite{pines_nozieres_book,giuliani_vignale_book}:
\begin{equation}\label{eq:Boltzmann}
\begin{split}
&\left[\partial_t+\bm v (\bm r,\bm p, t) \cdot \nabla_{\bm r}+\bm F(\bm r, \bm p,t)\cdot \nabla_{\bm p}\right]  f^{(1)}(\bm r,\bm p,t)= S^{\rm el}\{f^{(1)}(\bm r,\bm p',t)\}(\bm r,\bm p,t)+S^{\rm ee}\{f^{(1)}(\bm r,\bm p',t)\}(\bm r,\bm p,t),
\end{split}
\end{equation}
where $\bm v({\bm r}, {\bm p}, t) \equiv \nabla_{\bm p} {\cal H}(\bm r,\bm p,t)$ is the quasiparticle velocity, $\bm F(\bm r, \bm p,t)\equiv-\nabla_{\bm r}{\cal H}(\bm r,\bm p,t)=-e\bm E(\bm r,t)-\nabla_{\bm r}U_{\rm L}(\bm r, \bm p, t)  $ is the total force acting on quasiparticles, $\bm E(\bm r,t)=-\nabla_{\bm r}\phi(\bm r,t)$ being the electric field, $S^{\rm el}\{f^{(1)}(\bm r,\bm p',t)\}(\bm r,\bm p,t)$ is the collision integral that takes into account collisions with the lattice (i.e.~electron-phonon scattering) and electron-impurity collisions, while $S^{\rm ee}\{f^{(1)}(\bm r,\bm p',t)\}(\bm r,\bm p,t)$ is the collision integral for e-e scattering. 

To simplify Eq.~(\ref{eq:Boltzmann}) we introduce the following Ansatz\cite{pellegrino_prb_2017}
\begin{equation}\label{eq:f1Ansatz}
f^{(1)}(\bm r,\bm p,t)=f_0(\epsilon_{\bm p}^*)-f_0'(\epsilon_{\bm p}^*)\sum_{m=-\infty}^{+\infty}\mathcal{F}_m(\bm r, t)e^{im\theta_{\bm p}},
\end{equation}
where $f_0(\epsilon)=\{\exp[(\epsilon-\bar{\mu})/(k_{\rm B} T)]+1\}^{-1}$ is the equilibrium Fermi-Dirac distribution function at chemical potential $\bar{\mu}$ and temperature $T$, $f_0'(\epsilon)$ is its derivative with respect to the energy $\epsilon$. 

Inserting this ansatz in Eq.~(\ref{eq:Boltzmann}), retaining only terms that are linear in the coefficients $\mathcal{F}_m(\bm r,\theta_{\bm p},t)$, integrating over the energy $\epsilon_{\bm p}^*$, Fourier transforming with respect to time, and making use of the parametrization 
\begin{equation}
\frac{f_{\bm p\uparrow,\bm p'\uparrow}+f_{\bm p\uparrow,\bm p'\downarrow}}{2}=\frac{1}{L^2\mathcal{N}^*}\sum_{l=-\infty}^\infty F^{\rm s}_{|l|}e^{il(\theta_{\bm p}-\theta_{\bm p'})}
\end{equation}
of the Landau interaction function in terms of the so-called dimensionless Landau parameters~\cite{giuliani_vignale_book,note_landau_parameters} $F^{\rm s}_l$---where, $\mathcal{N}^*$ is the renormalized density of states at the Fermi level and $\theta_{\bm p}$ is the polar angle of the vector $\bm p$---we obtain: 
\begin{widetext}
\begin{equation}\label{eq:BoltzmannAnsatz}
-i\omega \sum_{m=-\infty}^{+\infty}\mathcal{F}_m(\bm r, \omega)e^{im\theta_{\bm p}}+ v_{\rm F}^* \hat{\bm p}\cdot \left[ \sum_{m=-\infty}^{+\infty}(1+F^{\rm s}_{|m|})\nabla\mathcal{F}_m(\bm r, \omega)e^{im\theta_{\bm p}}+e\bm E(\bm r,\omega)\right]
=-\sum_{m=-\infty}^{+\infty}\left[\Gamma^{\rm el}_m+\Gamma^{\rm ee}_m \right]\mathcal{F}_m(\bm r,\omega)e^{im\theta_{\bm p}}.
\end{equation}
\end{widetext}
Here,  $\bm E(\bm r,\omega)$ is the total electric field, i.e.~the sum of the external field and the field generated by the electron distribution itself (the Hartree self-consistent field), $v_{\rm F}^*\equiv |\nabla_{\bm p} \epsilon^*_{\bm p}|_{p=\hbar k_{\rm F}}$ is the Fermi velocity as renormalized by e-e interactions, $\hat{\bm p}=\bm p/|\bm p|$, and the relaxation coefficients $\Gamma^{\rm ee/el}_m$ are defined in terms of the respective linearized collision integrals by
\begin{equation}\label{eq:gammadefinition}
\Gamma_{m}^{\rm ee/el}=\int_{-\infty}^\infty d\epsilon_{\bm p}\int \frac{d\theta_{\bm p}}{2\pi}e^{-im \theta_{\bm p}}S^{\rm ee/el}_1\{-f'(\epsilon_{\bm p'})e^{i m\theta_{\bm p'}}\}(\bm p).
\end{equation}
Conservation of the particle number in collisions forces $\Gamma_0$ to vanish for all scattering processes.
Similarly, the conservation of total momentum forces $\Gamma_{\pm 1}$ to vanish for e-e collisions, while electron-lattice and electron-impurity processes are not subject to this constraint. 

We are now interested in solving Eq.~(\ref{eq:BoltzmannAnsatz}) in the presence of translational invariance.
To this aim, we perform a Fourier transform on the spatial variable, multiply (\ref{eq:BoltzmannAnsatz}) by $\exp[-in\theta_{\bm p}]$, and average over the angle $\theta_{\bm p}$.
This yields the infinite matrix equation 
\begin{widetext}
\begin{equation}
\label{eq:matrixequation}
\begin{pmatrix}
\ddots & \vdots & \vdots & \vdots & \vdots & \vdots & \reflectbox{$\ddots$} \\
\cdots & a_{-2} & b_{-1} & 0 & 0 & 0 & \cdots \\ 
\cdots & b^*_{-2} & a_{-1} & b_0 & 0 & 0 & \cdots \\ 
\cdots & 0 & b^*_{-1} & a_0 & b_1 & 0 & \cdots \\
\cdots & 0 & 0 & b^*_0 & a_1 & b_2 & \cdots \\ 
\cdots & 0 & 0 & 0 & b^*_1 & a_2 & \cdots \\
\reflectbox{$\ddots$} & \vdots & \vdots & \vdots & \vdots & \vdots & \ddots
\end{pmatrix}
\begin{pmatrix}
\vdots \\ \mathcal{F}_{-2} (\bm q,\omega)\\ \mathcal{F}_{-1}(\bm q,\omega) \\ \mathcal{F}_0(\bm q,\omega) \\ \mathcal{F}_1(\bm q,\omega) \\ \mathcal{F}_2 (\bm q,\omega)\\ \vdots
\end{pmatrix}
 = -\frac{iev_{\rm F}^*}{2} 
\begin{pmatrix}
\vdots \\ 0 \\ E^{(+)}(\bm q,\omega) \\ 0 \\E^{(-)}(\bm q,\omega) \\ 0 \\ \vdots
\end{pmatrix}~,
\end{equation}
\end{widetext}
where
\begin{equation}
a_n=\omega+i\Gamma_n^{\rm ee}+i\Gamma_n^{\rm el}~,
\end{equation}
and
\begin{equation}
b_n=b(1+F^{\rm S}_{|n|})~.
\end{equation}
Here, $b=-v_{\rm F}^*q^{(+)}/2$, with $q^{(\pm)}=q_x\pm i q_y$, and $E^{(\pm)}(\bm q,\omega)=E_x(\bm q,\omega)\pm iE_y(\bm q,\omega)$.

The solution of Eq.~(\ref{eq:matrixequation}) requires the inversion of the tridiagonal matrix $M$ appearing on the left hand side of this equation. In what follows we evaluate the relevant elements of $M^{-1}$, using the continued fraction method~\cite{lorentzen_book}, with the aim of calculating the response of the electron density to a longitudinal electric field.

Up to now our model has been completely general. For the purpose of obtaining a simple expression for the response function we make the following assumptions:
i) The electron-lattice and electron-impurity processes are characterized by only one parameter, i.e.~$\gamma$. We therefore have $\Gamma^{\rm el}_0=0$ and $\Gamma^{\rm el}_m=\gamma$ for $|m|>1$.
ii) The e-e collisions are described by only one parameter, i.e.~$\gamma_{\rm ee}$, resulting in $\Gamma^{\rm ee}_0$, $\Gamma^{\rm ee}_{\pm 1}=0$, and $\Gamma^{\rm ee}_m=\gamma_{\rm ee}$ for $|m|>1$.
iii) We consider only the zeroth- and first-order Landau parameters, $F^{\rm s}_0$ and $F^{\rm s}_1$, respectively, while all the $F^{\rm s}_l$ with $l\geq 2$ are set to zero. The solution method presented in the next Section can, however, be trivially generalized to any finite number of Landau parameters and relaxation rates.

With the aforementioned approximations, we find $a_0=\omega$, $a_{\pm 1}=\omega+i\gamma$, $a_{m}=\omega+i\gamma_{\rm tot}$, with $\gamma_{\rm tot}\equiv \gamma+\gamma_{\rm ee}$ for $|m|\geq 2$, while $b_0=b(1+F_0^{\rm s})$, $b_{\pm 1}=b(1+F_1^{\rm s})$ , and $b_m=b$ for $|m|\geq 2$. In the spirit of Ref.~\onlinecite{pellegrino_prb_2017}, we can obtain the hydrodynamic equations by truncating the matrix $M_{ij}$, retaining only the elements with $i, j\leq 2$.

For $i=0$ we obtain the continuity equation
\begin{equation}
-i\omega n(\bm q,\omega)+i\bm q \cdot \bm J(\bm q,\omega)=0~,
\end{equation}
where the induced density is directly related to $\mathcal{F}_0(\bm q,\omega)$ by
\begin{equation} \label{eq:Density}
n(\bm q,\omega )
\equiv\int d \bm p \,\delta f^{(1)}(\bm q,\bm p,\omega)
\approx{\cal N}^* \mathcal{F}_0(\bm q,\omega)~,
\end{equation}
and the current is related to $\mathcal{F}_{\pm 1}(\bm q,\omega)$ by
\begin{equation}\label{eq:Current}
\begin{split}
\bm J(\bm r,\omega)
 & \equiv\int d \bm p \,\bm v(\bm r,\bm p,\bm \omega)f^{(1)}(\bm r,\bm p,\omega)\approx\frac{{\cal N^*}v_{\rm F}^*}{2} (1+F_1^{\rm s})
\begin{pmatrix} \mathcal{F}_{-1}(\bm r,\omega)+\mathcal{F}_1(\bm r,\omega)\\
i\mathcal{F}_1(\bm r,\omega)-i\mathcal{F}_{-1}(\bm r,\omega)\end{pmatrix}~.
\end{split}
\end{equation}
In the above equations we neglected, for consistency, terms of higher order in the coefficients $\mathcal{F}_{n}(\bm q,\omega)$ and ignored the thermal smearing of the Fermi-Dirac function.

From the two equations for $i=\pm 1$ and using the equations for $i=\pm 2$ to eliminate $\mathcal{F}_{\pm 2}(\bm q,\omega)$, we obtain
\begin{equation}
-i\omega \bm J(\bm q,\omega)=-\gamma \bm J(\bm q,\omega)-\frac{(v_{\rm F}^*)^2(1+F_0^{\rm s})(1+F_1^{\rm s})}{2}i \bm q n(\bm q,\omega)-e\frac{(v_{\rm F}^*)^2\mathcal{N}^*(1+F_1^{\rm s})}{2}{\bm E}(\bm q,\omega)-\frac{(v_{\rm F}^*)^2(1+F_1^{\rm s})}{4(\gamma+\gamma_{\rm ee}-i\omega)}q^2\bm J(\bm q,\omega)~.
\end{equation}
By taking the limit $\omega \ll \gamma+\gamma_{\rm ee}$ in the last term and identifying
\begin{align}
\frac{(v_{\rm F}^*)^2(1+F_1^{\rm s})}{4(\gamma+\gamma_{\rm ee})} & =\nu^*,\\
\frac{v_{\rm F}^2 \mathcal{N}}{2} & =\frac{\bar{n}}{m},\\
\frac{v_{\rm F}^*}{v_{\rm F}}(1+F_1^{\rm s}) & =\frac{\mathcal{D}^*}{\mathcal{D}}, \label{eq:drudesub}\\
\frac{\bar{n}mv_{\rm F}v_{\rm F}^*(1+F_0^{\rm s})}{2} & =\frac{1}{K^*},\label{eq:compsub}\\
\end{align}
we obtain the Navier-Stokes equation (1) in the main text.
\section{Longitudinal response}
\label{app:densityresponse}

In this Section we calculate the density response to a longitudinal field. In this case, we write $\bm E(\bm q,\omega)=-i\bm q\phi(\bm q, \omega)$, yielding $E^{(\pm)}(\bm q,\omega)=-iq^{(\pm)}\phi(\bm q, \omega)$.
We are interested in calculating the density response, which, as stated in Eq.~(\ref{eq:Density}), is proportional to $\mathcal{F}_0(\bm q,\omega)$.
The proper density-density response function~\cite{giuliani_vignale_book} of the system is then given by
\begin{equation}\label{eq:chinn}
\begin{split}
\tilde{\chi}_{nn}(\bm q,\omega) & =\frac{\mathcal{N}^*\mathcal{F}_0(\bm q,\omega)}{-e\phi(\bm q,\omega)}
=-\mathcal{N}^*\left\{b\left[M^{-1}\right]_{0,-1}+b^*\left[M^{-1}\right]_{0,1} \right\}
=\mathcal{N}^*\frac{a_0[M^{-1}]_{00}-1}{1+F_0^{\rm s}}~.
\end{split}
\end{equation}
In writing the last equality we had to invert the matrix $M$ in Eq.~(\ref{eq:matrixequation}).
We also used: i) the Kramers rule expression for the inverse matrix elements $\left[M^{-1}\right]_{0,\pm 1} =- D_{\pm 1,0}/D$,  $[M^{-1}]_{00}=D_{0,0}/D$, where $D_{i,j}$ is the determinant of the matrix obtained from $M$ by suppressing the $i$-th row and the $j$-th column and $D=\det[M]$.
ii) The Laplace expansion on the $0$-th column of the determinant $D$, which yields $D= a_0D_{0,0}-b_0 D_{-1,0} -b_0^*D_{1,0}$.
iii) $b_0=b(1+F_0^{\rm s})$.

For a tridiagonal matrix $M$ in the form~(\ref{eq:matrixequation}), a diagonal element of the inverse matrix $M^{-1}$ can be expressed as a continued fraction\cite{lorentzen_book} 
\begin{equation}\label{eq:continuedfraction}
\begin{split}
[M^{-1}]_{00} & = \dfrac{1}{a_0 - \dfrac{b_1b_0^*}{a_1 - \dfrac{b_2b_1^*}{a_2 - \cdots}} - \dfrac{b_0b_{-1}^*}{a_{-1} - \dfrac{b_{-1}b_{-2}^*}{a_{-2} - \cdots}}}
=\dfrac{1}{a_0-\dfrac{2 |b|^2(1+F_0^{\rm s})(1+F_1^{\rm s})}{a_1+(1+F_1^{\rm s})\xi(q,\omega)}}~,
\end{split}
\end{equation}
where $\xi(q,\omega)$ respects the self-consistent equation
\begin{equation}\label{eq:xisc}
\xi(q,\omega) =\dfrac{-|b|^2}{a_2 - \dfrac{|b|^2}{a_2 -\dfrac{|b|^2}{ \cdots}}} = - \dfrac{|b|^2}{a_2 + \xi(q,\omega)}~.
\end{equation}
Solving for $\xi(q,\omega)$ and substituting the values of $a_2$ and $b$ we obtain
\begin{equation}\label{eq:xi}
\xi(q,\omega) = \frac{\omega+i\gamma_{\rm tot}}{2} \left[\sqrt{1-\frac{(v_{\rm F}^*)^2q^2}{(\omega+i\gamma_{\rm tot})^2}}-1\right]~.
\end{equation}
Here, we chose the solution of Eq.~(\ref{eq:xisc}) with the positive sign of the square root to make sure that the first-order expansion of $\xi$ in Eq.~(\ref{eq:xi}) in $|b|^2$ coincides with the truncation of the continued fraction up to first order in Eq.~(\ref{eq:xisc}). 

Making use of Eqs.~(\ref{eq:chinn})-(\ref{eq:continuedfraction}), and~(\ref{eq:xi}), we obtain the final result
\begin{widetext}
\begin{equation}\label{eq:chinn_final}
\begin{split}
& \tilde{\chi}_{nn}({\bm q},\omega) = \frac{(1+F^{\rm s}_1)(v_{\rm F}^*)^2\mathcal{N}^*q^2}{\omega(\omega + i\gamma)(1-F^{\rm s}_1)-(1+F^{\rm s}_1)\left[i\gamma_{\rm ee}\omega -\omega \sqrt{(\omega + i\gamma_{\rm tot} )^2 - (v_{\rm F}^*)^2q^2} + (v_{\rm F}^*)^2q^2(1+F^{\rm s}_0)\right]}~.
\end{split}
\end{equation}
\end{widetext}
Eq.~(\ref{eq:chinn_final}) is the semi-classical density-density response function of a 2D electron liquid, taking into account momentum-conserving and momentum-non-conserving collisions, and many-body effects through the renormalization of $v_{\rm F}$ and $\mathcal{N}_0$, and the Landau parameters  $F^{\rm s}_{0}$ and $F^{\rm s}_{1}$. 
This result can be easily converted into the longitudinal conductivity in Eq.~(2) of the main text using $\tilde{\chi}_{nn}(q,\omega)=-iq^2\sigma_{\rm L}(q,\omega)/(e^2\omega)$, Eqs.~(\ref{eq:drudesub})-(\ref{eq:compsub}), and $\mathcal{N}/\mathcal{N^*}=v_{\rm F}^*/v_{\rm F}$.
\section{Local capacitance approximation}
\label{app:capacitance}
The interaction potential between two electrons in a 2D system is 
\begin{equation}\label{eq:interactionpotential}
v_q=e^2G(q,0,0)~,
\end{equation}
where $G(q,z,z')$ is the electrostatic Green function satisfying 
\begin{equation}
q^2\epsilon_\parallel(z)G(q,z,z')-\partial_z[\epsilon_\perp(z)\partial_zG(q,z,z')]=4\pi \delta(z-z')~,
\end{equation}
where $z=0$ is the plane where electrons roam, and $\epsilon_{\parallel}$ ($\epsilon_{\perp}$) is the in-plane (out-of plane) dielectric constant of the dielectric environment. This equation must be supplemented by the boundary conditions at the metallic gate, i.e.
\begin{equation}
\frac{qG(q, z = -d^+, z')}{\epsilon_{\perp}(z = -d^+)\partial_z G(q, z = -d^+, z')} = \mathcal{Z}~.
\end{equation}
Here, $\mathcal{Z}$ is the dimensionless impedance of the metallic gate ($\mathcal{Z} = 0$ for a perfect conductor). In presence of screening by nearby conductors, the electrostatic Green function converges to a finite limit in the long-wavelength limit. 
It is therefore meaningful to define a capacitance per unit area
\begin{equation}
C\equiv \lim_{q \to 0}\frac{1}{G(q,0,0)}.
\end{equation}
\section{Comparison of the screening parameter in Single-layer and Bilayer Graphene}
\label{app:screening}
\begin{figure}[h!!]
\begin{overpic}[scale=1]{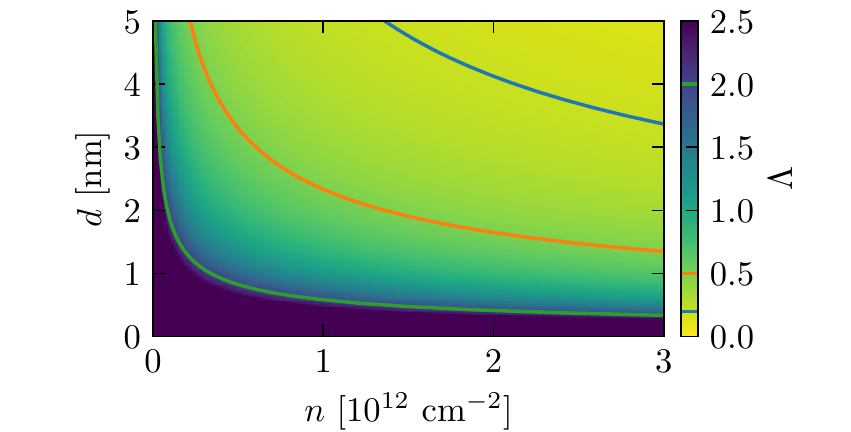}\put(2,115){(a)}
\end{overpic}
\begin{overpic}[scale=1]{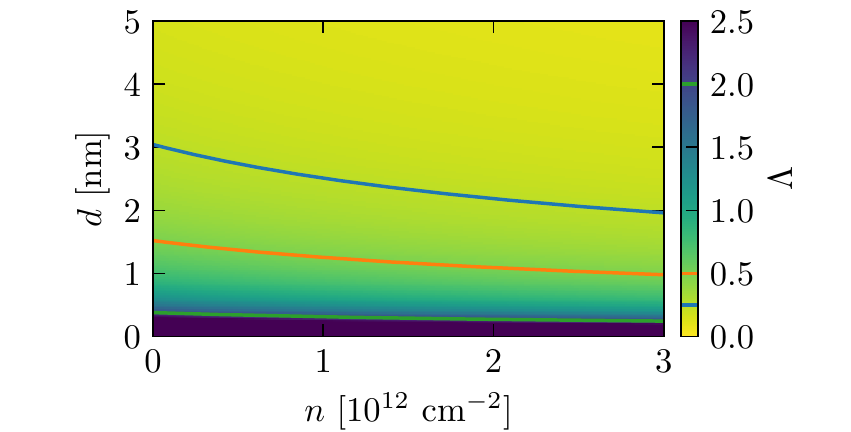}\put(2,115){(b)}
\end{overpic}\\
\caption{\label{fig:screening}
(a) Screening parameter $\Lambda$ as a function of electronic density $n$ and spacer thickness $d$ for a single-layer graphene/hBN/metal heterostructure like the one used in Ref.~\onlinecite{lundeberg_science_2017}. 
Results in this figure have been obtained by setting $\bar{\epsilon}_{zz}=3.5$ and $\mathcal{Z}=0$. Contour lines have been drawn for
$\Lambda=0.25$ (blue), $\Lambda=0.5$ (orange), and $\Lambda=0.25$ (green).
(b) Same as in panel (a) but for bilayer graphene. 
}
\end{figure}

The dimensionless quantity $\Lambda$ introduced in Eq.~(5) of the main text is the most important parameter of our theory since its value determines whether or not the crossover between the collisionless and hydrodynamic regimes is clearly discernible or not.
In Fig.~\ref{fig:screening} we show its value as a function of density and gate distance for an heterostructure made of graphene separated from a metal gate by an hBN slab of thickness $d$.
It is evident that single-layer graphene allows to reach larger values of $\Lambda$ with respect to bilayer graphene thanks to its smaller effective mass, especially at low densities. 
\section{Exact solution of plasmon equation}
\label{app:solution}
This Section is devoted to the calculation of the solutions of the plasmon equation $\epsilon_{\rm L}(q,\omega)=0$ with $\epsilon_{\rm L}(q,\omega)$ given by Eq.~(4) of the main text. 
By rearranging the various terms, the equation $\epsilon_{\rm L}(q,\omega)=0$ is equivalent to
\begin{equation}\label{eq:plmodes}
\begin{split}
&\sqrt{(\omega+i\gamma+i\gamma_{\rm ee})^2-(qv_{\rm F}^*)^2}=
\frac{(\Lambda^{-1}+\frac{K}{K^*})v_{\rm F}^*}{\omega v_{\rm F}}(qv_{\rm F})^2-\frac{2\mathcal{D}v_{\rm F}^*(\omega+i\gamma)}{\mathcal{D}^*v_{\rm F}}+(\omega+i\gamma+i\gamma_{\rm ee})~.
\end{split}
\end{equation}
Under the condition
\begin{equation}
{\rm Re}\left[\frac{(\Lambda^{-1}+\frac{K}{K^*})v_{\rm F}^*}{\omega v_{\rm F}}(qv_{\rm F})^2-\frac{2\mathcal{D}v_{\rm F}^*(\omega+i\gamma)}{\mathcal{D}^*v_{\rm F}}+(\omega+i\gamma+i\gamma_{\rm ee})\right]\geq 0~,
\end{equation}
this becomes
\begin{equation}\label{eq:plmodesfinal}
\left(\frac{qv_{\rm F}}{\omega}\right)^4+Q\left(\frac{qv_{\rm F}}{\omega}\right)^2+R=0~,
\end{equation}
with
\begin{equation}
Q = \frac{\omega\frac{\mathcal{D}^*}{\mathcal{D}}\frac{v_{\rm F}^*}{v_{\rm F}} -2\left[(2\frac{v_{\rm F}^*}{v_{\rm F}}-\frac{\mathcal{D}^*}{\mathcal{D}})(\omega+i\gamma)-i\frac{\mathcal{D}^*}{\mathcal{D}}\gamma_{\rm ee}\right](\Lambda^{-1}+\frac{K}{K^*})}{\omega(\Lambda^{-1}+\frac{K}{K^*})^2\frac{\mathcal{D}^*}{\mathcal{D}}\frac{v_{\rm F}^*}{v_{\rm F}}}
\end{equation}
and
\begin{equation}
R =-\frac{4(\omega+i\gamma)[\frac{\mathcal{D}^*}{\mathcal{D}}(\omega + i \gamma+i\gamma_{\rm ee}) -\frac{v_{\rm F}^*}{v_{\rm F}}(\omega+i \gamma)]}{\omega^2\frac{v_{\rm F}^*}{v_{\rm F}}(\frac{\mathcal{D}^*}{\mathcal{D}})^2(\Lambda^{-1}+\frac{K}{K^*})^2}~.
\end{equation}
Eq.~(\ref{eq:plmodesfinal}) is a quadratic equation for $(qv_{\rm F}^*/\omega)^2$ with solutions
\begin{equation}\label{eq:qpl}
\left(\frac{qv_{\rm F}}{\omega}\right)^2=-\frac{Q}{2}-\frac{\sqrt{Q^2-4R}}{2}=-Q\frac{1+\sqrt{1+4RQ^{-2}}}{2}~,
\end{equation}
where we discarded the second solution since it gives ${\rm Im}(q) < 0$, which has no physical meaning. 
Finally, Eq.~(\ref{eq:qpl}) is equivalent to Eq.~(6) of the main text if we define the following quantities as the velocity and damping, respectively:
\begin{equation}\label{eq:velocity}
S_\omega=\frac{v_{\rm F}}{\sqrt{{\rm Re}\left[\left(\frac{qv_{\rm F}}{\omega}\right)^2\right]}}~,
\end{equation}
and
\begin{equation}\label{eq:damping}
\Gamma_\omega=\frac{\omega}{2} \frac{{\rm Im}\left[\left(\frac{qv_{\rm F}}{\omega}\right)^2\right]}{{\rm Re}\left[\left(\frac{qv_{\rm F}}{\omega}\right)^2\right]}~.
\end{equation}
In order to obtain expressions for $S_{\omega}$ and $\Gamma_{\omega}$ in the collisionless limit, we expand up to linear order in $\gamma/\omega$ and $\gamma_{\rm ee}/\omega$. We find $Q\approx Q_0+i\gamma/\omega Q_\gamma+i\gamma_{\rm ee}/\omega Q_{\rm ee}$ and 
$R\approx R_0+i\gamma/\omega R_\gamma+i\gamma_{\rm ee}/\omega R_{\rm ee}$, where
\begin{align}
Q_0 & =-\frac{2(2\frac{v_{\rm F}^*}{v_{\rm F}}-\frac{\mathcal{D}^*}{\mathcal{D}})(\Lambda^{-1}+\frac{K}{K^*})-\frac{\mathcal{D}^*}{\mathcal{D}}\frac{v_{\rm F}^*}{v_{\rm F}} }{(\Lambda^{-1}+\frac{K}{K^*})^2\frac{\mathcal{D}^*}{\mathcal{D}}\frac{v_{\rm F}^*}{v_{\rm F}}}~,\\
Q_\gamma & =-\frac{2(2\frac{v_{\rm F}^*}{v_{\rm F}}-\frac{\mathcal{D}^*}{\mathcal{D}})}{(\Lambda^{-1}+\frac{K}{K^*})\frac{\mathcal{D}^*}{\mathcal{D}}\frac{v_{\rm F}^*}{v_{\rm F}}}~,\\
Q_{\rm ee} & =\frac{2\frac{\mathcal{D}^*}{\mathcal{D}}}{(\Lambda^{-1}+\frac{K}{K^*})\frac{\mathcal{D}^*}{\mathcal{D}}\frac{v_{\rm F}^*}{v_{\rm F}}}~,\\
R_0 & =-\frac{4(\frac{\mathcal{D}^*}{\mathcal{D}}-\frac{v_{\rm F}^*}{v_{\rm F}})}{\frac{v_{\rm F}^*}{v_{\rm F}}(\frac{\mathcal{D}^*}{\mathcal{D}})^2(\Lambda^{-1}+\frac{K}{K^*})^2}~,\\
R_\gamma & =-\frac{8(\frac{\mathcal{D}^*}{\mathcal{D}}-\frac{v_{\rm F}^*}{v_{\rm F}})}{\frac{v_{\rm F}^*}{v_{\rm F}}(\frac{\mathcal{D}^*}{\mathcal{D}})^2(\Lambda^{-1}+\frac{K}{K^*})^2}~,\\
R_{\rm ee} & =-\frac{4}{\frac{v_{\rm F}^*}{v_{\rm F}}\frac{\mathcal{D}^*}{\mathcal{D}}(\Lambda^{-1}+\frac{K}{K^*})^2}~.
\end{align}
The replacement of the approximate expressions for $Q$ and $R$, with the coefficients above, into Eq.~(\ref{eq:qpl}) results in the following expression
\begin{equation}\label{eq:qpcoll}
\left(\frac{qv_{\rm F}}{\omega}\right)^2\approx\left[-Q_0\frac{1+\sqrt{1+4R_0Q_0^{-2}}}{2}\right]+\frac{i\gamma}{\omega}\frac{Q_\gamma \left[-Q_0\frac{1+\sqrt{1+4R_0Q_0^{-2}}}{2}\right]+R_\gamma}{Q_0\sqrt{1+4R_0Q_0^{-2}}}+\frac{i\gamma_{\rm ee}}{\omega}\frac{Q_{\rm ee} \left[-Q_0\frac{1+\sqrt{1+4R_0Q_0^{-2}}}{2}\right]+R_{\rm ee}}{Q_0\sqrt{1+4R_0Q_0^{-2}}}~.
\end{equation}
Finally, the replacement of Eq.~(\ref{eq:qpcoll}) into Eqs.~(\ref{eq:velocity}) and~(\ref{eq:damping}) yields Eqs.~(7) and (8) of the main text.

On the other hand, we can obtain the corresponding results in the hydrodynamic limit by expanding (\ref{eq:qpl}) for $\gamma_{\rm ee} \gg \omega$. This leads to
\begin{equation}\label{eq:qphydro}
\left(\frac{qv_{\rm F}}{\omega}\right)^2 + R^{\rm h} = 0~,
\end{equation}
with
\begin{equation}
R^{\rm h} = \frac{-2(\omega+i\gamma)}{\frac{\mathcal{D}^*}{\mathcal{D}}\omega\left[(\Lambda^{-1}+\frac{K}{K^*}) -i\frac{v_{\rm F}^*\omega}{2v_{\rm F}(\gamma+\gamma_{\rm ee})}\right]}~.
\end{equation}
Replacing Eq.~(\ref{eq:qphydro}) into Eqs.~(\ref{eq:velocity}) and~(\ref{eq:damping}) results in Eqs.~(9) and (10) of the main text.

\section{Coupling to a near-field probe}
\label{app:coupling}
We consider an external dipole source of strength $p$ and frequency $\omega$, positioned at height $z_d$ and with its axis pointing in the $\hat{\bm z}$ direction.
This can be described by an oscillating charge density
\begin{equation}
\rho_{\rm ext}(\bm q,z,\omega)=-p \delta'(z-z_d)~,
\end{equation}
where $\delta'(z)$ is the derivative of $\delta(z)$ with respect to its argument. The field it generates is $\bm E_{\rm d}(\bm r,z,\omega)=-\nabla\phi_{\rm d}(\bm r,z,\omega)$, with
\begin{equation}\label{eq:dipolepotential}
\phi_{\rm d}(\bm r,z,\omega)=\int\frac{d\bm q}{(2\pi)^2}e^{i \bm q \cdot \bm r}\int dz' G(q,z,z')\rho_{\rm ext}(\bm q,z',\omega)=\int\frac{d\bm q}{(2\pi)^2}
e^{i \bm q \cdot \bm r}pG'( q,z_d,z)~,
\end{equation}
where $G(q,z,z')$ is the electrostatic Green function defined in Sect.~III and $G'(q,z,z')\equiv \partial_zG(q,z,z')$.

The field $\bm E_{\rm d}(\bm r,z,\omega)$ induces a charge oscillation in the electron liquid, which absorbs an average power 
\begin{equation}
\begin{split}
\langle W\rangle (z_d)& =\int d^2\bm r \frac{1}{2}{\rm Re}[-e\bm J^*(\bm r,\omega)\cdot \bm E_{\rm d}(\bm r,0,\omega)]\\
& = \frac{1}{2}{\rm Re}\left[i\omega e\int d^2\bm r \, n^*(\bm r,\omega)\phi_{\rm d}(\bm r,0,\omega)\right]\\
& = \frac{1}{2}{\rm Re}\left[i\omega e\int\frac{ d^2\bm q}{(2\pi)^2} \, n^*(\bm q,\omega)\phi_{\rm d}(\bm q,0,\omega)\right]\\
& = \frac{1}{2}{\rm Re}\left[-i\omega e^2\int\frac{ d^2\bm q}{(2\pi)^2} \, \chi_{nn}^*(q,\omega)|\phi_{\rm d}(\bm q,0,\omega)|^2\right]\\
& = -\frac{\omega e^2}{2}\int\frac{ d^2\bm q}{(2\pi)^2} \, |\phi_{\rm d}(\bm q,0,\omega)|^2{\rm Im }\left[\chi_{nn}(q,\omega)\right]
 = \frac{\omega e^2}{2}\int\frac{ d^2\bm q}{(2\pi)^2} \, |\phi_{\rm d}(\bm q,0,\omega)|^2{\rm Im}\left[\frac{-1}{v_q\epsilon_{\rm L}(q,\omega)}\right]\\
& = \frac{\omega e^2p^2}{2}\int\frac{ d^2\bm q}{(2\pi)^2} \, \frac{|G'(q,z_d,0)|^2}{G'( q,0,0)}\mathcal{L}(q,\omega)
 = \frac{\omega e^2p^2}{2}\int\frac{ d q q}{2\pi} \, \frac{[G'(q,z_d,0)]^2}{G'(q,0,0)}\mathcal{L}(q,\omega)~.
\end{split}
\end{equation}
In the second line we used integration by parts and the continuity equation, in the third Parseval's theorem, in the fourth $n(\bm q,\omega)=\chi_{nn}(q,\omega)(-e)\phi_{\rm d}(\bm q,0,\omega)$, $\chi_{nn}(q,\omega)$ being the density-density response function of the electron system,
in the fifth ${\rm Im}\left[\chi_{nn}(q,\omega)\right]= {\rm Im}\left[1/(v_q\epsilon_{\rm L}(q,\omega))\right]$, and, in the sixth, we defined the loss function $\mathcal{L}(q,\omega)=-{\rm Im}\left[1/\epsilon_{\rm L}(q,\omega)\right]$, and made use of the definition of the interaction potential (\ref{eq:interactionpotential}), and of the Fourier transform of (\ref{eq:dipolepotential}). 

Since we are interested only in the power fed into the AP, which will be denoted by the symbol $\langle W\rangle_{\rm AP}(z_d)$, we consider only the contribution to the above integral coming from wave vectors smaller than the edge $q_{\rm eh}$ of the intra-band electron-hole continuum, $q_{\rm eh}(\omega)\equiv\lim_{\Lambda \to \infty}{\rm Re}[q_{\rm p}(\omega)]$.

We finally define the coupling efficiency as the ratio  $\eta_{z_d}(\omega)\equiv\langle W\rangle_{\rm AP}(z_d)/W_{\rm Larmor}$ between the power fed into the AP  and that radiated by the dipole if it was in free space, given by Larmor's formula
\begin{equation}
W_{\rm Larmor}=\frac{p^2\omega^4}{3c^3}~.
\end{equation}
The final result reads as following
\begin{equation}\label{eq:etadefinition}
\eta_{z}(\omega)=\frac{ 3c^3}{4\pi\omega^3}\int_0^{q_{\rm eh}(\omega)} dq q \frac{[G'(q,z,0)]^2 }{G(q,0,0)}\mathcal{L}(q,\omega)~.
\end{equation}
The previous formula has been used to produce the numerical results in Fig.~3 of the main text.

In the case of a 2D material separated from a perfect metal located at $z=-d$ by a dielectric spacer, we can approximate, for long wavelengths, $G(q,z,0)\approx e^{-qz}/C$ if $z>0$ and $G(q,z,0)\approx (z+d)/(dC)$ $0>z>-d$. Furthermore, if dissipation is small, we can approximate the loss function, in the relevant range of wave vectors, as a delta function peak $\mathcal{L}(q,\omega)\approx \pi |Z|{\rm Re}(q_{\rm p})\delta(q-{\rm Re}(q_{\rm p}))$ with $Z\equiv [{\rm Re}(q_{\rm p}) \partial_q\epsilon_{\rm L}(q,\omega)|_{q=q_{\rm p}}]^{-1}=-[2+q_P \partial_q\sigma_{\rm L}(q,\omega)|_{q=q_{\rm p}}/\sigma_{\rm L}(q_{\rm p},\omega)]^{-1}\approx -1/2$.
Using these approximations in Eq.~(\ref{eq:etadefinition}) we get the approximate result in Eq.~(10) of the main text.